\documentclass[%
 reprint,
 amsmath,amssymb,
 aps,
]{revtex4-2}

\usepackage{graphicx}
\usepackage{dcolumn}
\usepackage{bm}

\begin{document}

\preprint{APS/123-QED}

\title{Negative temperature coefficient of resistivity due to the itinerant spin fluctuations in metallic V$_{0.3}$Ti$_{0.7}$ alloy}

\author{SK Ramjan}%
 \affiliation{Free Electron Laser Utilization Laboratory, Raja Ramanna Centre for Advanced Technology, Indore 452 013, India}
 \affiliation{Homi Bhabha National Institute, Training School Complex, Anushakti Nagar, Mumbai 400 094, India}
 \author{Asi Khandelwal}%
 \affiliation{Free Electron Laser Utilization Laboratory, Raja Ramanna Centre for Advanced Technology, Indore 452 013, India}
 \affiliation{Homi Bhabha National Institute, Training School Complex, Anushakti Nagar, Mumbai 400 094, India}
\author{L. S. Sharath Chandra}
\affiliation{Free Electron Laser Utilization Laboratory, Raja Ramanna Centre for Advanced Technology, Indore 452 013, India}
 \affiliation{Homi Bhabha National Institute, Training School Complex, Anushakti Nagar, Mumbai 400 094, India}
 \email{lsschandra@rrcat.gov.in}
\author{M. K. Chattopadhyay}
\affiliation{Free Electron Laser Utilization Laboratory, Raja Ramanna Centre for Advanced Technology, Indore 452 013, India}
 \affiliation{Homi Bhabha National Institute, Training School Complex, Anushakti Nagar, Mumbai 400 094, India}

\date{\today}

\begin{abstract}
Few concentrated disordered binary metallic alloys show a negative temperature coefficient of resistivity (TCR), which is quite unusual. V$_{0.3}$Ti$_{0.7}$ is one such alloy that shows resistivity exceeding 100 $\mu\Omega$ cm and exhibits negative TCR. The addition of ferromagnetic rare-earth Gd, which is insoluble in the body-centered cubic V-Ti matrix, changes the negative TCR to positive (when Gd conc. $\geq$ 2 at.$\%$). Evidence of spin polarization of conduction electrons of the V-Ti matrix by Gd clusters is obtained from the magnetization experiments as well as from the anomalous component of the Hall effect. Our analysis suggests that the additional scattering due to the distribution in the electron-spin fluctuation interaction arising from the substitution of titanium in vanadium is the origin of the negative TCR. The Gd clusters polarize the conduction electrons, leading to the suppression of spin fluctuations, resulting in the positive TCR.
\end{abstract}

\maketitle


\section{\label{sec:level1}Introduction}

When a random disorder is introduced into a crystalline medium through add-atoms or vacancies, it results in a random distribution of atomic potentials \cite{mott}. The scattering of electrons from these randomly distributed potentials can reduce the electron mean free path $l_e$ below the inter-atomic distances $a$ \cite{mott}. This leads to a negative temperature coefficient of resistivity (TCR) with a resistivity $\rho$ = 100-150~$\mu \Omega$-cm. Such situations are observed when two dissimilar metals, such as Cu-Zr, Ti-Al, V-Al, Ni-Cr, Mn-Pd etc., are used to form alloys where the potential energy of the two different atoms in an alloy is significantly different \cite{moo73}. However, similar analogy to the negative TCR in the body centered cubic (bcc) $\beta$-V$_{1-x}$Ti$_x$ alloys\cite{sab19, mat14, ike90, sas90, isi85, pre75, col74, ros74} is surprising as V and Ti atoms lie adjacent to each other in the periodic table.

Alternative explanations were explored to account for the negative TCR in the $\beta$-V$_{1-x}$Ti$_x$ alloys. Collings pointed out that the amount of athermal $\omega$-phase increases with increasing $x$ in the $\beta$-V$_{1-x}$Ti$_x$ alloys \cite{col74}. By comparing the electrical resistivity of titanium binary alloys with their microstructure, the strength of the negative TCR of the $\beta$-V$_{1-x}$Ti$_x$ alloys was attributed to the amount of $\omega$-phase present. However, our recent studies showed that the resistivity of certain V$_{0.6}$Ti$_{0.4}$ samples did not increase in spite of an increase in the $\omega$-phase \cite{ram24, asi24}. On the other hand, Prikul and coworkers\cite{pre75,ros74} argued quantitatively by comparing the electrical resistivity of $\beta$-V$_{1-x}$Ti$_x$ alloys with the localized spin fluctuation theory given by Rivier and Zlatic \cite{riv72} that the negative TCR in $\beta$-V$_{1-x}$Ti$_x$ alloys is due to the localized spin fluctuations. Again, our studies showed that the spin fluctuations present in the  $\beta$-V$_{1-x}$Ti$_x$ alloys are itinerant in nature.\cite{mat14} Formation of a pseudo gap for a small amount of transition metals in titanium was also proposed to account for the negative TCR in titanium alloys \cite{shc83}. However, the compositions studied belong to $\alpha$ or $\alpha$-rich $\alpha$+$\beta$ phases of the Ti-transition metal binaries. The density of states calculations does show any pseudo gap in the $\beta$-V$_{1-x}$Ti$_x$ alloys.\cite{mat14} Therefore, negative TCR in the $\beta$-V$_{1-x}$Ti$_x$ alloys is not due to the pseudo gap formation. Weak-localization was also considered by Sasaki and Moto \cite{sas90} in certain quenched V$_{1-x}$Ti$_x$ alloys to explain the $\sqrt{T}$ and $\sqrt{H}$ dependence of resistivity at low temperatures. However, after annealing these alloys still maintain a negative TCR, but without $\sqrt{T}$ and $\sqrt{H}$ dependence of resistivity. Therefore, the intrinsic negative TCR in the $\beta$-V$_{1-x}$Ti$_x$ alloys is not due to weak-localization effects.

Recently, we have shown that the addition of gadolinium to the $\beta$-V$_{0.6}$Ti$_{0.4}$ alloy decreases the residual resistivity in spite of increase in the disorder \cite{sab21a}. Note that Gd clusters precipitate along the grain boundaries and retain their elemental characteristic with ferromagnetic transition at 295~K \cite{sab21}. We argued that the reduction in the residual resistivity in these alloys is due to the suppression of itinerant spin fluctuations (ISF) present in the $\beta$-phase, by the polarization of the conduction electrons around and along the grain boundaries \cite{sab21a}. This prompted us to revisit negative TCR in the $\beta$-V$_{1-x}$Ti$_x$ alloys in the realm of ISF.    

Apart from this, the $\beta$-V$_{1-x}$Ti$_{x}$ alloy superconductors are also considered to be promising for high field applications especially in neutron environment.\cite{ram24, sab21, bar20, nag20, ayy18, tai07, nod04} These alloys show several other interesting properties such as phonon dominated thermal conductivity in the superconducting state in contrary to the normal state where electrons carry a significant amount of total heat \cite{sab19}, high field paramagnetic effect \cite{mat13, mat16, ram22}, phonon softening and strong influence of spin fluctuations \cite{mat14, mat14a} etc. On the technological front, the $\beta$-V$_{1-x}$Ti$_x$ superconducting alloys has a critical current density ($J_C$) of about 100~Amm$^{-2}$ or less \cite{mat15}. By forming multi-filamentary wires, the $J_C$ of V$_{1-x}$Ti$_x$ alloys has been improved to 700~Amm$^{-2}$ \cite{tai07, tak08}. Successive cold rolling and annealing the V$_{1-x}$Ti$_x$ wires prepared through diffusion reaction method improved $J_C$ further to 850~Amm$^{-2}$ \cite{tak10}. Recently, we have shown that by adding rare earth elements (RE) in small quantities to V$_{0.60}$Ti$_{0.40}$ alloy improves the $J_C$ to about 1000~Amm$^{-2}$ or less \cite{ram24}. Further, successive cold rolling and annealing at 450~$^0$C, the $J_C$ of RE added V$_{0.60}$Ti$_{0.40}$ alloys improved to 2000-8000~Amm$^{-2}$ in absence of magnetic field and about 500~Amm$^{-2}$ in 7~T \cite{ram24}. 

In this article, we show that the TCR changes from negative to positive by the addition of gadolinium to the $\beta$-V$_{0.30}$Ti$_{0.70}$ alloy. The onset of positive TCR in these alloys is also associated with the appearance of anomalous Hall effect along with the long range ferromagnetism below 295~K. Our analysis suggests that these observed properties are due to the suppression of ISF by the polarization of the conduction electrons of the $\beta$-V$_{0.30}$Ti$_{0.70}$ phase. We also show that the width of the superconducting transition temperature decreases with increasing Gd content in $\beta$-V$_{0.30}$Ti$_{0.70}$ alloy. With these observations, we argue that the negative TCR in $\beta$-V$_{1-x}$Ti$_{x}$ alloys is due to the local variations in the electron-spin fluctuation coupling constant at the atomic scale.      

\section{\label{sec:level1}Experimental Details}

The polycrystalline V$_{0.3-x}$Gd$_x$Ti$_{0.7}$ ($x$ = 0, 0.01, 0.02, 0.035) and V$_{0.3-y}$La$_y$Ti$_{0.7}$ ($y$ = 0.01, 0.02) samples were synthesized by arc melting 99.99\% vanadium and 99.99\% titanium with either 99.9\% gadolinium or 99.9\% lanthanum in 99.999\% pure argon atmosphere. The resulting ingots were flipped and re-melted 5-6 times to ensure homogeneity. Electrical resistivity and Hall effect measurements were performed using a Physical Property Measurement System (PPMS, Quantum Design, USA). The Hall resistivity was measured using the 5-probe geometry and was also averaged over positive and negative magnetic field sweeps to eliminate any contribution from longitudinal resistivity. DC magnetization was measured using a Superconducting Quantum Interference Device based Vibrating Sample Magnetometer (MPMS-3 SQUID-VSM, Quantum Design, USA). In this article, we use the nomenclature Gd$Z$ or La$Z$ to represent the samples where $Z$ is the amount of Gd/La (atomic percent) in V$_{0.3-x}$Gd$_x$Ti$_{0.7}$ alloy.     

\section{\label{sec:level1}Results and Discussion}

\begin{figure}[htb]
    \centering
    \includegraphics[width=\columnwidth]{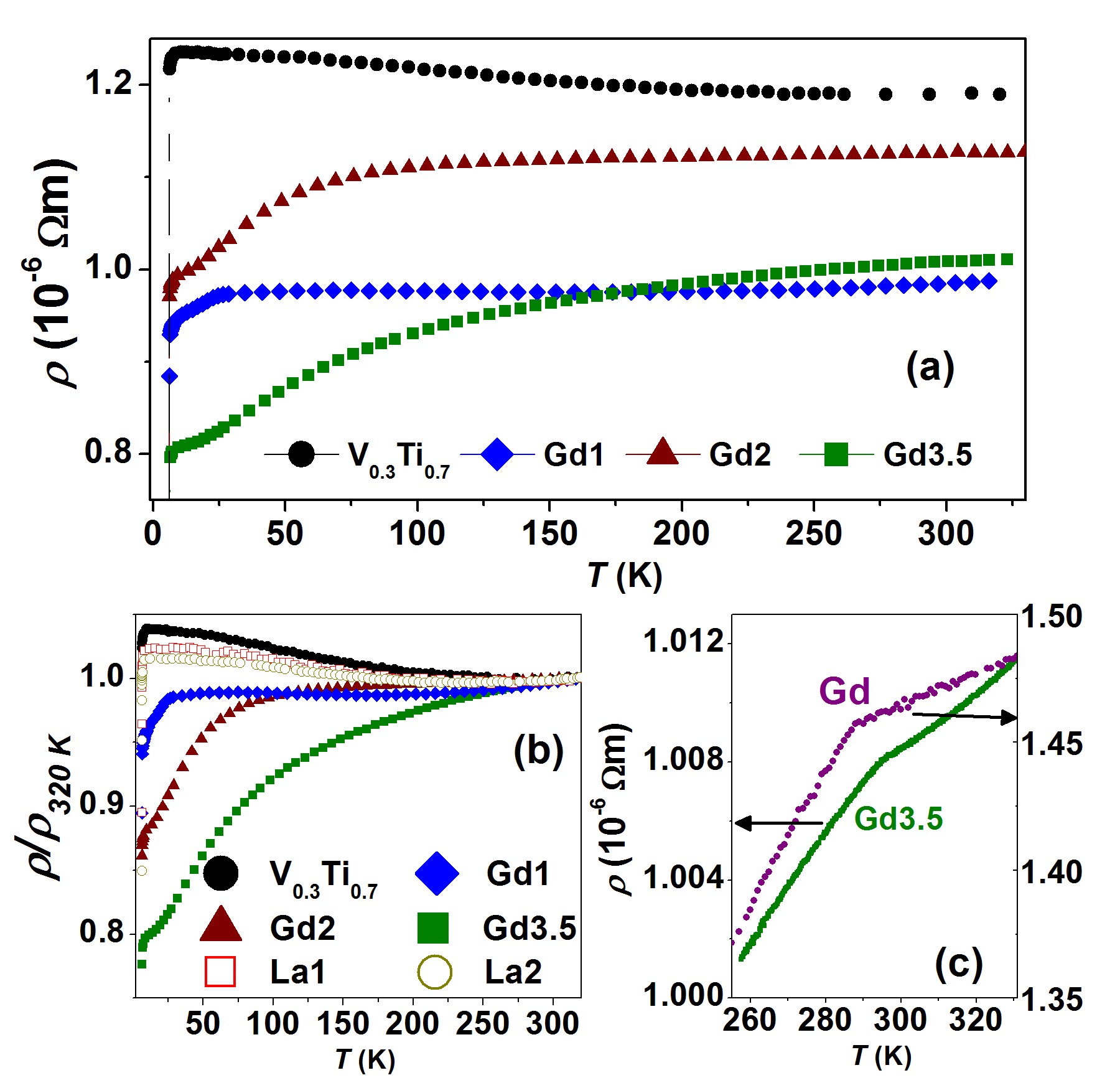}
    \caption{(a) Temperature dependence of zero field resistivity in the V$_{0.3}$Ti$_{0.7}$-Gd alloys. (b) Same as (a) but normalized by resistivity value at 320 K for V$_{0.3}$Ti$_{0.7}$-RE alloys. The temperature coefficient of resistivity changes from negative to positive with Gd addition. (c) Change of slope in resistivity at the paramagnetic to ferromagnetic transition temperature in pure Gd and Gd3.5 alloy.}
    \label{4}
\end{figure}

Gadolinium-rich hexagonally close packed (hcp) phase precipitates along the grain boundaries of the $\beta$-V$_{1-x}$Ti$_x$ alloys when Gd is added in small quantities, leading to the reduction in the grain size of the $\beta$-matrix.\cite{ram24, sab21}  In the present alloys, the average size of the Gd precipitates are 0.69 $\pm$ 0.48~$\mu$m in Gd1, 1.1 $\pm$ 0.6~$\mu$m in Gd2, and 1.12 $\pm$ 0.6~$\mu$m in Gd3.5. The temperature dependence of resistivity ($\rho$($T$)) of V$_{0.3}$Ti$_{0.7}$-Gd alloys is shown in Fig. \ref{4}(a). The V$_{0.3}$Ti$_{0.7}$ alloy has a negative TCR \cite{sab19} with a $\rho_0$ = $\rho(8~K)$ of 123~$\mu \Omega$ cm. The mean free path ($l_e$) of the V$_{0.3}$Ti$_{0.7}$ alloy estimated from the $\rho_0$ is smaller than the interatomic distance.\cite{sab19} 

Addition of Gd to the V$_{1-x}$Ti$_x$ alloys reduces the overall resistivity indicating that the spin disorder scattering is the major contribution to the resistivity in V$_{1-x}$Ti$_x$ alloys. Resistivity of the Gd1 alloy starts to drop with decreasing temperature below 70~K, giving rise to positive TCR.  In Gd2 and Gd3.5 alloys, the positive TCR is over entire temperature range of measurement. On the other hand, addition of La to V$_{0.30}$Ti$_{0.70}$ alloy does not result in positive TCR (Fig. \ref{4}(b)) at any temperature above the $T_C$. The $\rho$($T$) of La containing alloys mimic the parent V$_{0.30}$Ti$_{0.70}$ alloy. The $\rho$($T$) of the both elemental Gd and Gd3.5 alloy shows a slope change at around 295~K which match with the paramagentic to ferromagnetic transition of Gd (Fig. \ref{4}c).  

\begin{figure}
    \centering
    \includegraphics[width=\columnwidth]{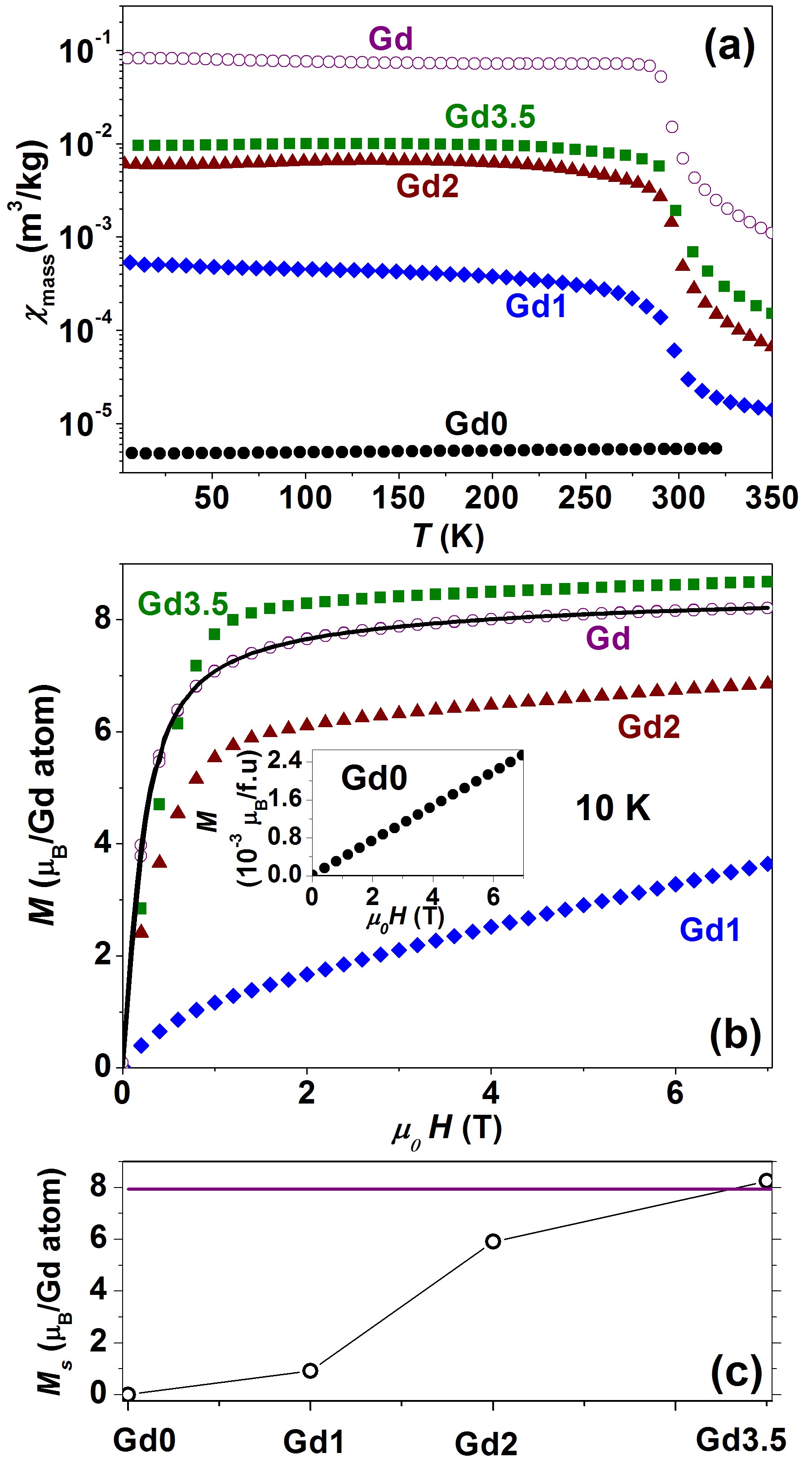}
    \caption{(a) Temperature dependent magnetization of the V$_{0.3}$Ti$_{0.7}$-Gd alloys indicating the paramagnetic to ferromagnetic transition at 295~K and the absence of the same in V$_{0.3}$Ti$_{0.7}$. For reference, the magnetization curve of elemental Gd is shown. (b) Field dependence of magnetization at 10 K for the V$_{0.3}$Ti$_{0.7}$-Gd alloys. The insert to (b) shows the variation of saturation magnetization with Gd content.}
    \label{2}
\end{figure}

\begin{figure*}[!htp]
     \centering
    \includegraphics[width=\textwidth]{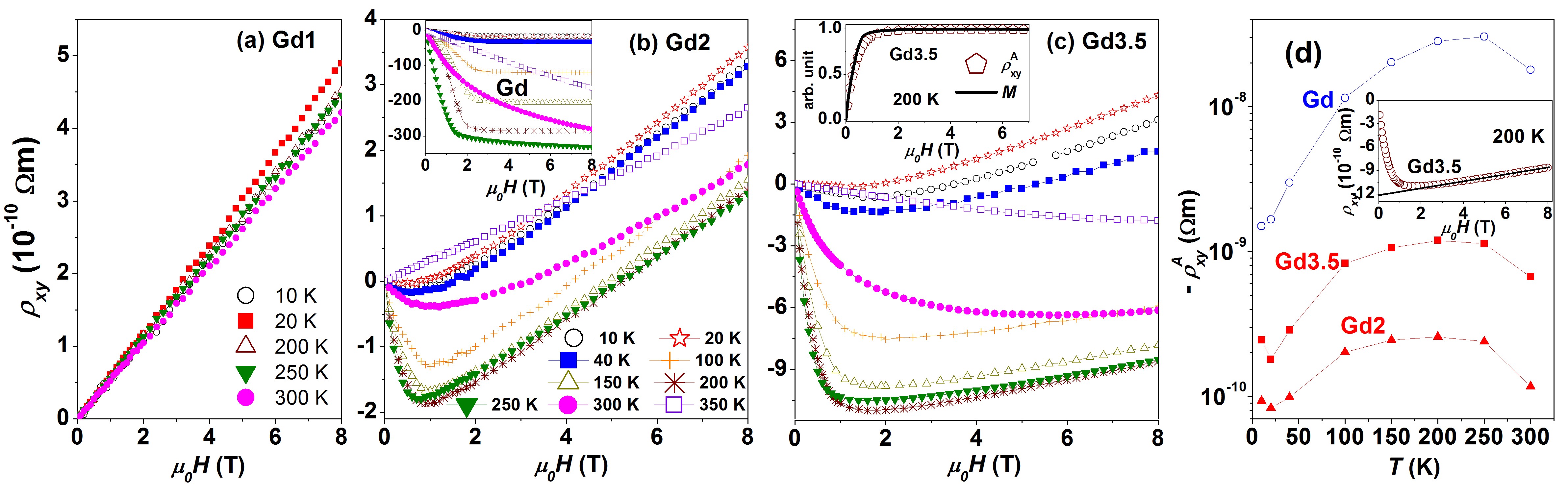}
    \caption{Field dependence of hall resistivity at various temperatures for (a)Gd1, (b) Gd2 and (c) Gd3.5 alloys. The inset to (b) shows the same for Gd. The inset to (c) shows the magnetization and anomalous hall resistivity (each normalized to their maximum value) as a function of magnetic field at 200 K. (d) Temperature dependence of anomalous hall resistivity for Gd2 and Gd5 alloys. The same for elemental Gd has been shown for comparison. The inset shows the method used to extract the anomalous contribution to hall resistivity.}
    \label{3}
\end{figure*}

The temperature dependence of field cooled DC susceptibility ($\chi$) (figure \ref{2}(a)) of the Gd-containing alloys exhibit a paramagnetic to ferromagnetic transition at $T_{curie}$ = 295~K, a characteristic feature of elemental Gd. This indicates that Gd clusters retain their magnetism due to the localization of $f$-electrons within the Gd atom. In Gd1, $\chi$ increases with decreasing temperature below $T_{curie}$. For the alloys with higher Gd content, the $\chi$ is nearly constant below $T_{curie}$. Figure \ref{2}(b) shows the field dependence of magnetization of the V$_{0.3}$Ti$_{0.7}$-Gd alloys along with the elemental Gd. The symbols represent the experimental data points. Although the Gd1 alloy exhibits a tendency of magnetic saturation in low fields, magnetization continues to rise linearly at high fields. The tendency of saturation is much stronger for the alloys with a higher Gd content; however, the rise of magnetization at high fields is less significant. The inset to the Fig. \ref{2}(b) shows the field dependence of magnetization of the V$_{0.3}$Ti$_{0.7}$ parent alloy at 10~K which has no tendency of magnetic saturation. The black solid line in the e main n panel of Fig. \ref{2}(b) is constructed by adding the experimentally measured magnetization of V$_{0.3}$Ti$_{0.7}$ with that of the elemental Gd in same molar ratios as those in V$_{0.3}$Ti$_{0.7}$-Gd alloys. Interestingly, the constructed curves for all the compositions coincide over one another and match with the experimental data of elemental Gd as M(V$_{0.3}$Ti$_{0.7}$) $<<$ M(Gd). On the other hand, the magnetization per Gd atom of Gd1 and Gd2 are quite lower than that of elemental Gd, it is higher for the Gd3.5 alloy for fields higher than {\bf 1~T}. These results indicate a partial screening of magnetic moments of Gd clusters by the conduction electrons of the $\beta$-matrix. When the number of clusters increase due to increasing Gd content in the system, the polarized conduction electrons (which participate in screening) from different clusters start interacting with each other. This leads to a long-range magnetic ordering which is manifested as magnetic saturation in the Gd2 alloy at high fields. With further increase in the Gd content (as in the case of Gd3.5 alloy), polarization of conduction electrons increases to an extent that the total magnetic moment of the system now exceeds that of the elemental Gd. The technical saturation magnetization ($M_{s}$) measured by extrapolating of Arrott plots for Gd1, Gd2 and Gd3.5 alloys turns out to be 0.92 $\mu_{B}$/Gd atom, 5.91 $\mu_{B}$/Gd atom and 8.25 $\mu_{B}$/Gd atom, respectively (Fig. \ref{2}(c)). The non-linear increase in $M_{s}$ with Gd concentration is also an indication of a coalescence like phenomena, when  Gd clusters start to interact through the conduction electrons of the V-Ti matrix as the concentration of Gd increases. Notably, the net effective moment observed in the Gd3.5 alloy exceeds the value expected for pure Gd (7.98 $\mu_B$,\cite{blundell} marked by straight line in Fig. \ref{2}(c)).

Both the resistivity and the magnetization of the V$_{0.3-x}$Gd$_x$Ti$_{0.7}$ alloys indicate the polarization of the conduction electrons, suggesting that the negative TCR in V$_{1-x}$Ti$_x$ alloys is related to the spin fluctuations. To prove this, we present the Hall effect in the V$_{0.3-x}$Gd$_x$Ti$_{0.7}$ alloys in Fig. \ref{3}. The Hall resistivity ($\rho_{xy}$) of Gd1 alloy (Fig.\ref{3}(a)) and the parent Gd0 alloy (not shown here) varies linearly with the applied field at all temperatures measured between 10-300~K. The positive slope of $\rho_{xy}$ vs. $H$ in Gd0 indicates that the holes are the major carriers, which is in agreement with the report available in literature \cite{bandyopadhyay1986}. The $\rho_{xy}$($H$) of Gd2 and Gd3.5 (Fig \ref{3}(b) and \ref{3}(c)) rise rapidly (with negative slope) in low fields followed by a linear increase (with positive slope) in higher fields. These features clearly indicate the presence of anomalous Hall effect (AHE) in these alloys. The Hall resistivity is usually expressed as $\rho_{xy}=R_{0}\mu_{0}H + \mu_{0}R_{s}M$. Here, $\mu_{0}$ = magnetic permeability, $R_0$ = ordinary Hall coefficient, $R_s$ = anomalous Hall coefficient, $M$ = magnetization. The first term denotes the ordinary Hall resistivity ($\rho_{xy}^O$) which arises due to the Lorentz force. The second term denotes the anomalous Hall resistivity ($\rho_{xy}^A$) and depends on the magnetization. 
 
 The inset to figure \ref{3}(c) shows that the scaling of magnetization and the $\rho_{xy}^A$. The magnetization values ($M'$) used in the inset to figure \ref{3}(c) are obtained in the following way: $M'(H)=M(H)-aH$, where $a$ is the slope of $M(H)$ in fields above 3 T. Discrepancy between $\rho_{xy}^A(H)$ and $M(H)$ was observed when actual $M(H)$ values were plotted. {\bf  In general, the anomalous contribution is much larger than the ordinary, as is evident from the Hall resistivity data of Gd provided in the inset to figure \ref{3}(b).} The $\rho_{xy} (H)$ for Gd at various temperatures is shown in the inset to figure \ref{3}(b). The magnitude of Hall resistivity increases with increases in temperature up to 250 K, and then decreases with further increase in temperature. This is in agreement with reports in literature, where Hall resistivity is expected to peak around the ferromagnetic transition temperature \cite{baily2005}. The Hall resistivity of Gd remains negative at all temperatures and in all fields (up to 8 T). In contrast, the Hall resistivity changes its sign from negative to positive at some temperatures above certain field in the Gd2 and Gd3.5 alloys due to the positive and negative contribution from the ordinary and anomalous contributions to Hall resistivity, respectively. The anomalous Hall resistivity ($\rho^A_{xy}$) can be estimated by extrapolating the high field linear part back to zero, as shown in inset to figure \ref{3}(d) for Gd3.5 at 200 K. The temperature dependence of anomalous contribution to Hall resistivity is plotted for Gd2, Gd3.5 and Gd in figure \ref{3}(d). The $\rho_{xy}^A(T)$ for Gd2 and Gd3.5 follow the same trend which is roughly similar to that observed in the case of Gd. A slight upturn at low temperatures is found in Gd2 and Gd3.5, which is absent in Gd. Moreover, the peak in $\rho_{xy}^A(T)$ is observed around 250 K in case of Gd, where as the peak position shifts towards 200 K in case of Gd2 and Gd3.5 alloys. The slope of the linear part ($ = R_{0}$) multiplied by the magnetic field gives $\rho_{xy}^O$. In Gd2 at 200~K, the $\rho^A_{xy}$ = -2.55$\times 10^{-10}\ \Omega$m, which is smaller in magnitude than $\rho_{xy}^O$ ($H$ = 8 T) = 4$\times 10^{-10}\ \Omega$m. This results in a positive $\rho_{xy}$ in the highest field measured. However, in Gd3.5, the anomalous contribution grows ($\rho^A_{xy}$ = -1.75$\times 10^{-9}\ \Omega$m) and is greater than the magnitude of the ordinary Hall component even in the highest field ($\rho_{xy}^O$ ($H$ = 8 T) = 3.1$\times 10^{-10}\ \Omega$m). Hence, increase in Gd concentration significantly increases the anomalous contribution to Hall resistivity, without affecting much the ordinary Hall component. The ratio $\rho^A_{xy}$/$\rho_{xx}$ is often used to quantify the strength of AHE. This turns out to be 2.09 $\times 10^{-4}$ and 1.68 $\times 10^{-3}$ in Gd2 and Gd3.5, respectively. $\rho^A_{xy}$/$\rho_{xx}$ is of the order of 10$^{-4}$ in extrinsic dilute magnetic semiconductors (DMS), where AHE is due to the presence of an extrinsic mechanism. For instance, in Co-(La,Sr)TiO$_3$ the AHE is of the order of $10^{-4}$ \cite{zhang2007}. The nanosize magnetic clusters embedded in the non magnetic matrix polarises the nearby conduction electrons which results in net spin polarisation and gives rise to AHE. The clusters do not interact with each other due to their low density \cite{zhang2007} and hence, the contribution to the AHE is small. However, in intrinsic DMS and in Gd, $\rho^A_{xy}$/$\rho_{xx}$ is of the order of $10^{-2}$. In such cases, AHE is related to intrinsic mechanism. The magnitude of $\rho^A_{xy}$/$\rho_{xx}$ in Gd3.5 is almost one order higher than Gd2. Increase in the number of Gd clusters and the decrease in inter-Gd cluster distance, allow Gd clusters to interact with each other through the conduction electrons of the V-Ti matrix. This establishes long range ferromagnetic order in V$_{0.3}$Ti$_{0.7}$-Gd alloys, even when the Gd clusters are far apart. 

  \begin{figure} 
    \centering
    \includegraphics[width=\columnwidth]{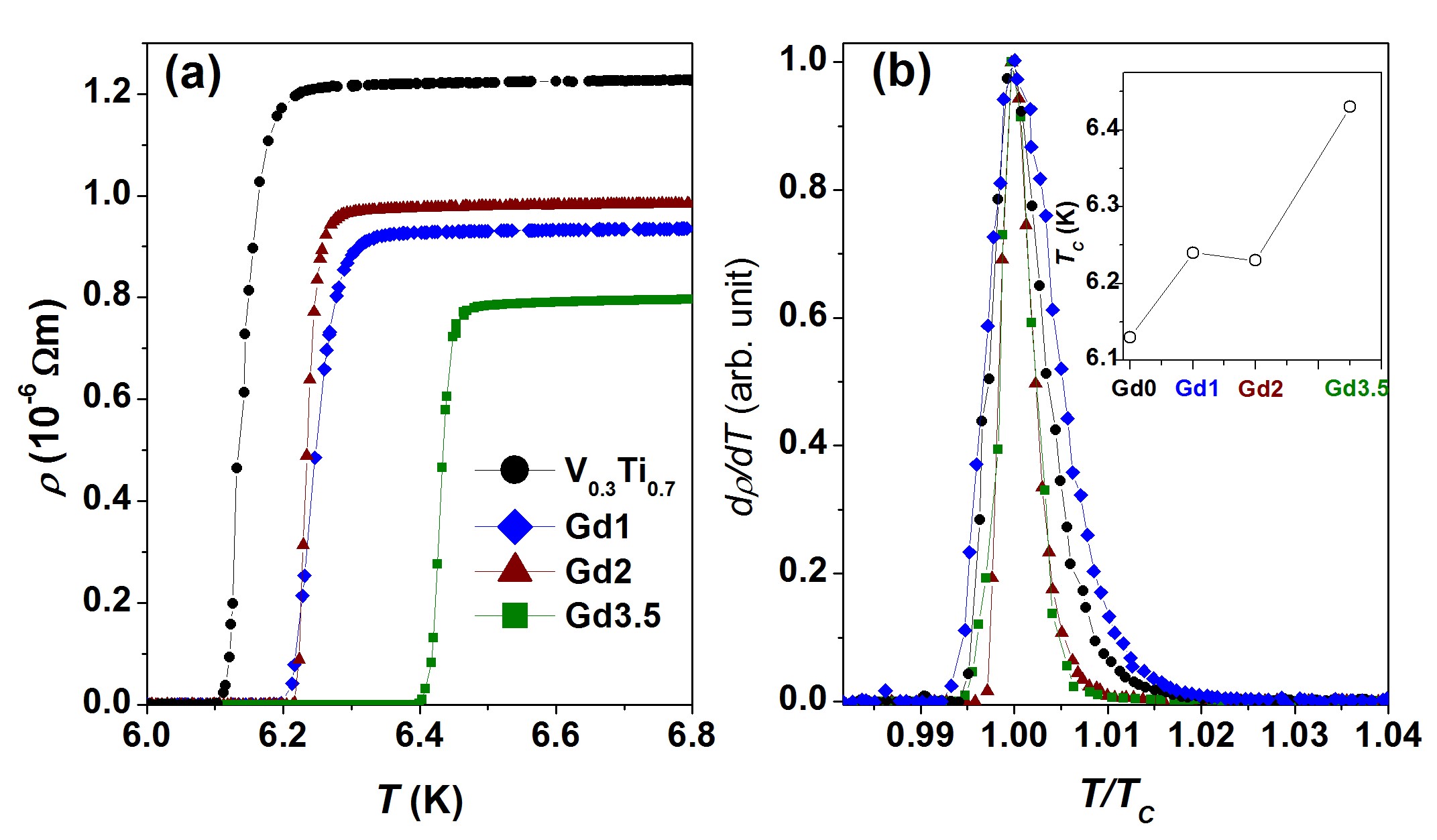}
    \caption{(a) Temperature dependence of resistivity near $T_C$ for V$_{0.3}$Ti$_{0.7}$-Gd alloys. (b) Temperature derivative of resistivity plotted against temperature normalised by $T_C$ for V$_{0.3}$Ti$_{0.7}$-Gd. FWHM of the peaks become narrower and the broadening at temperature above $T_C$ is reduced when Gd $\geq$ 2 at.\%.}
    \label{5}
\end{figure}

From the above analysis of temperature and magnetic field dependence of resistivity and Hall effect, it is clear that the suppression of ISF by the polarization of the conduction elections is responsible for the change of TCR from negative to positive. It is well known that magnetic field and disorder can suppress the spin fluctuations \cite{stritzker1979}. However, the magnetic field required to quench spin fluctuations will be quite high \cite{ikeda1984}, considering the very high spin fluctuation temperature ($T_{sf}$) in the V-Ti alloys \cite{mat14}. The strong magnetic moments of Gd clusters can effectively increase the internal magnetic field and suppress the spin fluctuations. Gd addition has also been found to suppress the ferromagnetic SF in the UAl$_2$ and UAl$_3$ through the exchange interaction between the localized Gd 4\textit{f} moments and conduction electrons \cite{oliveira1993, lucaci1996}. From these arguments, it appears that the ISF is responsible for the negative TCR in $\beta$-V$_{1-x}$Ti$_x$ alloys. However, ISF is maximum in vanadium and reduces with the addition of titanium \cite{mat14}. Similar contradiction appeared in the width of the superconducting transition temperature and fluctuation conductivity \cite{mat14a}. We argued at that time that the appearance of significant fluctuation conductivity at higher Ti concentration is due to the local distribution in the electron-spin fluctuation coupling constant ($\lambda_{sf}$) \cite{mat14a}. We argue here, that this local distribution of $\lambda_{sf}$ is also responsible for the negative TCR. The local distribution of $\lambda_{sf}$ generates additional spin disorder scattering. By the addition of Gd, the suppression of the spin fluctuations reduces the local distribution of $\lambda_{sf}$. As a proof, we provide the temperature dependence of resistivity ($d\rho/dT$) and its derivative across the $T_C$ in figure \ref{5}. We observe that (i) the $T_C$ increases from 6.12 K in Gd0 to 6.45 K in Gd3.5 (Fig. \ref{5}(a)) despite the onset of long range ferromagnetism by the addition of Gd. (ii) decrease in the width of the $T_C$ as seen in the $d\rho/dT$ (Fig. \ref{5}(b)) with the increase in the Gd amount in the V$_{0.3}$Ti$_{0.7}$-Gd alloy. Both these results suggest the reduction of incoherent scattering due to the local distribution of $\lambda_{sf}$ by the addition of Gd into V$_{0.3}$Ti$_{0.7}$-Gd alloy which results in the positive TCR as well as improved superconducting properties.   

\section{Summary and conclusion}
In this paper, we studied the effect of Gd (1, 2, 3.5 at.\%) addition on the microstructure, magnetization, resistivity and Hall effect of the V$_{0.3}$Ti$_{0.7}$ alloy. From the observation of anomalous Hall effect and saturation magnetization in V$_{0.3}$Ti$_{0.7}$-Gd (for Gd concentration $\geq$ 2 at.\%), we have established that the presence of Gd-rich clusters in the non-magnetic V-Ti matrix can polarize the V-Ti conduction electrons. As a result, the spin fluctuations present in the V-Ti matrix is suppressed, thereby increasing the $T_C$ and reducing the width of normal to superconducting phase transition. Interestingly, the temperature coefficient of resistivity which is negative in the V$_{0.3}$Ti$_{0.7}$ alloy, changes to positive when Gd ($\ge$ 2 at.\%) is added. From the analysis of experimental observations, we claim that the spin fluctuations are the origin of negative TCR in the V$_{0.3}$Ti$_{0.7}$ alloy and addition of Gd suppresses the spin fluctuations and changes the TCR from negative to positive.


\nocite{*}

 \begin{thebibliography}{1}

\bibitem{ram24}
SK. Ramjan, A. Khandelwal, S. Paul, L. S. Sharath Chandra, R. Singh, R. Venkatesh, K. Kumar, R. Rawat, S. Dutt, A. Sagdeo, P. Ganesh and M. K. Chattopadhyay, Journal of Alloys and Compounds \textbf{976}, 173321 (2024).

\bibitem{sab21} 
S. Paul, Sk. Ramjan, R. Venkatesh, L. S. Sharath Chandra, and M. K. Chattopadhyay, IEEE Trans. Appl. Supercond. 31 (2021) 8000104. 

\bibitem{bar20}
P. J. Barron, A. W. Carruthers, J. W. Fellowes, N. G. Jones, H. Dawson, and E. J. Pickering, Scri. Mater. 176 (2020) 12-16.

\bibitem{nag20} 
T. Nagasaka, and T. Muroga, {\it Comprehensive Nuclear Materials} 6 (2020) 1-18.

\bibitem{ayy18} 
A. Ayyagari, R. Salloom, S. Muskeri, S. Mukherjee, Materialia 4 (2018) 99.

\bibitem{tai07} 
M. Tai, K. Inoue, A. Kikuchi, T. Takeuchi, T. Kiyoshi, Y. Hishinuma, IEEE Trans. Appl. Supercond. 17 (2007) 2542-2545.

\bibitem{nod04} 
T. Noda, T. Takeuchi, and M. Fujita,  J. Nucl. Mater. 329-333 (2004) 1590-1593.

\bibitem{sab19}
S. Paul, L. S. Sharath Chandra and M. K. Chattopadhyay, J. Phys. Condens. Matter {\bf 31}, 475801-1-8 (2019) 

\bibitem{mat14}
M. Matin, L. S. Sharath Chandra, S. K. Pandey, M. K. Chattopadhyay and S. B. Roy, The Eur. Phys. J. B \textbf{87}, 131-1-10 (2014).

\bibitem{ike90} M. Ikeda, S-ya Komatsu, T. Sugimoto, and K. Kamei, J. Jpn. Inst. Metals {\bf 54}, 743-751 (1990).

\bibitem{sas90} T. Sasaki, and Y. Muto, Physica B {\bf 165\&166} 291-292 (1990).

\bibitem{isi85} M. Isino, J. Jpn. Phys. Soc. {\bf 54}, 3848-3587 (1985). 

\bibitem{pre75} A. F. Prekul, V. A. Rassokhin, and N. V. Volkenshtein, JETP Lett. {\bf 22}, 209-210 (1975) [Pisma Zh. Eksp. Teor. Fiz. {\bf 22}, 433-436 (1975)].

\bibitem{col74} E. W. Colings, Phys. Rev. B {\bf 9} 3989-3999 (1974).

\bibitem{ros74} V. A. Rassokhin, and N. V. Volkenshtein, E. P. Romanov, and A. F. Prekul, Sov. Phys. JETP {\bf 39}, 166-168 (1974) [Zh. Eksp. Teor. Fiz. {\bf 66}, 348-353 (1974)].

\bibitem{mat13} Md. Matin, L. S. Sharath Chandra, M. K. Chattopadhyay, M. N. Singh, A. K. Sinha, and S. B. Roy, Supercond. Sci. Technol. {\bf 26}, 115005, (2013).

\bibitem{mat16} Md. Matin, M. K. Chattopadhyay, L. S. Sharath Chandra, and S. B. Roy, Supercond. Sci. Technol. {\bf 29}, 025003 (2016).

\bibitem{ram22} SK. Ramjan, L. S. Sharath Chandra, Rashmi Singh, and M. K. Chattopadhyay, Supercond. Sci. Technol. {\bf 35}, 105006 (2022). 

\bibitem{sab21a} S. Paul, SK. Ramjan, L. S. Sharath Chandra, and M. K. Chattopadhyay, Mater. Sci. Engg. B {\bf 274}, 115462 (2021). 

\bibitem{mat14a} Md. Matin, L. S. Sharath Chandra, R. Meena, M. K. Chattopadhyay, A. K. Sinha, M. N. Singh, and S. B. Roy, Physica B {\bf 436} 20-25 (2014).

\bibitem{mat15} Md. Matin, L. S. Sharath Chandra, M. K. Chattopadhyay, R. K. Meena, R. Kaul, M. N. Singh, A. K. Sinha, and S. B. Roy, Physica C: Supercond. Appl. {\bf 512}, 32 (2015).

\bibitem{tak08} T. Takeuchi, H. Takigawa, M. Nakagawa, N. Banno, K. Inoue, Y. Iijima, A. Kikuchi, Supercond. Sci. Technol. {\bf 21}, 025004 (2008).

\bibitem{tak10} T. Takeuchi, H. Takigawa,  N. Banno, M. Nakagawa, M. Iwatani, K. Inoue, Y. Hishinuma, A. NIshimura, AIP Conf. Proc. {\bf 1219}, 263-270 (2010).

\bibitem{mott} N. F. Mott, Metal-Insulator Transitions (Taylor \& Francis, London (1974)).

\bibitem{moo73} J. H. Mooij, Phys. Stat. Sol. A {\bf 17}, 521 (1973).

\bibitem{asi24} Asi Khandelwal, Nida Mirza, L. S. Sharath Chandra, R. Singh, A. Sagdeo, and M. K. Chattopadhyay, J. Supercond. Novel. Mag. (accepted) (2024). 

\bibitem{riv72} N. River, V. Zlatic, J. Phys. F: Metal Phys. {\bf 2}, L87 (1972).

\bibitem{shc83} A. S. Shcherbakov, A. F. Prekul, and R. V. Pomortsev, Phil. Mag. B {\bf 47}, 63-72 (1983)

\bibitem{mcl82} D. S. McLachlan, Solid State Commun. \textbf{42},  521-523 (1982)

\bibitem{blundell} S. Blundell, Magnetism in condensed matter, OUP Oxford (2001).

\bibitem{bandyopadhyay1986} B. Bandyopadhyay, P. Watson, Y. Bo, D. G. Naugle and V. M. Nicoli, Zeitschrift für Physik B Condensed Matter \textbf{63}, 207–211 (1986).

\bibitem{stritzker1979} B. Stritzker, Phys. Rev. Lett. \textbf{42}, 1769 (1979)

\bibitem{zhang2007} S. X. Zhang, W. Yu, S. B. Ogale, S. R. Shinde, D. C. Kundaliya, Wang-Kong Tse, S. Y. Young, J. S. Higgins, L. G. Salamanca-Riba, M. Herrera, L. F. Fu, N. D. Browning, R. L. Greene and T. Venkatesan, Physical Review B \textbf{76}, 085323 (2007).

\bibitem{baily2005} S. A. Baily and M. B. Salamon, Physical Review B \textbf{71}, 104407 (2005).

\bibitem{ikeda1984} K. Ikeda, K. A. Gschneidner Jr, R. J. Stierman, T. W. E. Tsang, and O. D. McMasters, Physical Review B \textbf{29}, 5039 (1984).

\bibitem{oliveira1993}
N. A. De Oliveira, A. A. Gomes and A. Troper, Hyperfine Interactions \textbf{80}, 1067-1070 (1993).

\bibitem{lucaci1996}
P. Lucaci, E. Burzao and I. Lupsa, Journal of Alloys Compounds \textbf{238}, L4-L6 (1996).


\end{thebibliography}

\end{document}